\RequirePackage{fix-cm}
\documentclass[smallextended]{svjour3}       
\smartqed  
\usepackage{graphicx}
\usepackage{soul}
\usepackage{natbib}
\usepackage{hyperref} 
\journalname{European Journal for Philosophy of Science}
\begin{document}

\title{On the interpretation of Feynman diagrams, or, did the LHC experiments observe $H\rightarrow\gamma\gamma$?}


\titlerunning{Feynman diagrams and Higgs discovery}        

\author{Oliver Passon\footnote{School of Mathematics and Natural Sciences, University of Wuppertal, 42119 Wuppertal, Germany}}

\authorrunning{xxx} 
\authorrunning{O. Passon} 

\institute{Oliver Passon \at
            University of Wuppertal\\
             Gau{\ss}str. 20,              42119 Wuppertal,              Germany\\
              \email{passon@uni-wuppertal.de}           
}

\date{Received: date / Accepted: date}

\maketitle

\begin{abstract}
According to the received view Feynman diagrams are a bookkeeping device in complex perturbative calculations. Thus, they do not provide a representation or model of the underlying physical process. This view is in apparent tension with scientific practice in high energy physics, which analyses its data in terms of ``channels''. For example the Higgs discovery was based on the observation of the decay $H\rightarrow\gamma\gamma$ -- a process which can be easily represented by the corresponding Feynman diagrams. I take issue with this   tension and show that on closer analysis the story of the Higgs discovery should be told differently.  
\end{abstract}


 \section{Introduction}

Probably most physicists (and philosophers of physics) would agree that Feynman diagrams do not represent or model the underlying physical process in any closer meaning of these terms. Instead, these diagrams are a calculational tool or ``bookkeeping device'' for perturbative calculations in quantum field theory. Expressed pointedly, they visualize {\em formulae} and not physical {\em processes}.  In this paper I will support this claim.  

However, common and successful scientific practice in high energy physics  appears to be in tension with this view. {Take as a prominent example the recent discovery of the Higgs boson. 
Here, according to the usual narrative, the Higgs was detected via its dominant decays, e.g.  in the $H\rightarrow ZZ \rightarrow 4\ell$ (here, $\ell$ denotes an electron or a muon) or $H\rightarrow\gamma\gamma$  channel. These channels are characterized by a set of  Feynman diagrams and the account of the production and decay processes is usually just a verbal description of the corresponding diagrams. If there really were such a close connection between the experimental measurement and single Feynman diagrams the literal reading of them would be strongly supported}.    

{Now, the goal of this paper is twofold. First I will show that this tension is only apparent, that is, the Higgs discovery (which will serve as our case study) does  not presuppose a realistic interpretation of Feynman diagrams. In turn this leads to our second claim, namely that the common parlance  is misleading and that the story of the Higgs discovery needs to be told differently. Expressed pointedly, the LHC experiments did not observe a process like $H\rightarrow\gamma\gamma$ since the very category (``observation'') is not applicable. In order to avoid any misunderstanding, I do not question the correctness of the results. {It is just that they do not support the common narrative which singles out individual amplitudes of the scattering matrix while the mathematical framework presupposes that {\em all} relevant contributions (and this could include irreducible background as well) are taken into account.} That is, the corresponding analyses of ATLAS and CMS provide clear evidence for standard model cross-sections involving Higgs terms with a corresponding mass $m_H\approx 125$ GeV.  In this more carefully specified sense it is certainly true that the Higgs has been discovered at LHC.}

In order to substantiate these claims I need to provide the basis for the debate. Sec.~\ref{int} is devoted to the interpretation of Feynman diagrams. Much of the debate in the literature deals with specific aspects, like the status of ``virtual particles''. {We will briefly survey the arguments from the literature, widen the scope and introduce new arguments  to strengthen the claim that Feynman diagrams have no representational or modeling function proper}. All this conflicts with their visual nature and intuitive appeal -- {two aspects which have contributed to the success of this diagrammatic method}. In fact,  the origin of this ongoing debate can be traced back to the introduction of the Feynman diagrams in 1948 and their subsequent formalization by Dyson.  In order to highlight this relation I will briefly indicate the historical roots of the debate in Sec.~\ref{history}.

While all this should be rather uncontroversial, I believe that the specific application to the scientific practice in high energy physics is missing in the literature on this subject. In Sec.~\ref{higgs} I will therefore investigate the Higgs discovery at LHC as a case study. As indicated above our goal is to demonstrate that contrary to  first impressions a literal reading of Feynman diagrams is not supported by these results. 

In closing this introductory section I should say a few words regarding  terminology and methodology. The reader may have noted that I do not distinguish   between the alleged ``representational'' and the ``modeling'' function of Feynman diagrams. These notions are closely related, since the success of a {\em model} may be based on its ability to {\em represent} certain features of the target system. 
That is, I apply the term ``model'' in a rather conventional sense. 
Following Meinhard \citet[p. 126]{kuhlmann2010} I will call the interpretation of Feynman diagrams ``realistic'' or ``literal'' when it is assumed that the elements of the diagrams can be mapped onto something in the outer physical world. Fortunately, one is  not committed to any specific form of mapping; the literature contains suggestions ranging from ``isomorphic'', ``partially isomorphic'' or even less formal requirements \citep{models2012}. I believe that this captures the usual usage within modeling sciences, i.e. here models are studied in order to generate claims and to learn something about the target system \citep[p. 51]{frigg2017}.\footnote{However, ``representation'' and ``models'' are complex issues and in general they should not be equated (for a recent overview on the debate see e.g. \cite{frigg2017}). There are many non-model-based forms of representation (e.g. measurements provide a representation  of processes in nature, or lexicographical representations such as words).} 

At the same time there are other approaches to the concept of model in which the representational role is not the only (or even primary) function. For example  Michael  \citet{stoeltzner2017a,stoeltzner2017b} has argued that Feynman diagrams should be viewed as models in the more sophisticated sense of ``models as mediators'' as suggested by Mary Morgan and Margaret Morrison \citep{morgan1999}. Here, models are granted a more autonomous function. On this view Feynman diagrams may have more complex  representational functions than the one indicated above. I believe that this is an important line of research, even more so if one assumes that a modeling function of Feynman diagrams in the conventional sense is excluded. Our objective is exactly to strengthen this premise.\footnote{Also, Adrian \cite{wuethrich2012} suggests a modeling function of Feynman diagrams which focuses on the  more coarse-grained aspects of the description. However, his presentation suffers from a rather incomplete list of arguments against a realistic reading of Feynman diagrams he tries to circumvent. I believe that against the background of the arguments advanced here his claim can not be sustained.}  

Letitia \citet{meynell2008} argues that Feynman diagrams do have a representational function. However, this claim is not based on any property of these diagrams which had been overlooked before. Instead, she weakens the   requirements for representation by separating ``representation'' from ``denotation''. By this move the epistemic function of Feynman diagrams is  left open  \citep[p. 56]{meynell2008}. While this is an interesting approach I believe that such a weaker notion of ``representation'' is at odds with its common understanding in science. In any event the point at issue in this paper is exactly the epistemic function of Feynman diagrams. 

Finally, all this is loosely related to the realism-debate. According to a popular myth ``representations'' are like a mirror image or a copy of the represented object, i.e. they are ``realistic representations'' \citep[p. 50]{frigg2017}. Roman Frigg and James Nguyen remark that there are non-realistic representations and that ``representation is a much broader notion than mirroring'' \citep[p. 50]{frigg2017}. I will sometimes talk about Feynman diagrams as understood ``literally'' or being taken to ``depict'' or to ``visualize'' the physical process. However, this wording is not intended to suggest the  narrow notion of ``representation as mirroring''. The question is not whether the representation is ``realistic'' but whether the  system represented is ``real''. Thus, this minimal form of realism is usually presupposed, namely the independent existence of the target system.   In this sense part of the debate about the interpretation of Feynman diagrams has been framed e.g. as the question ``Are virtual particles real?'' (compare e.g. \citet{weingard}).

\section{A brief history of  Feynman diagrams\label{history}}

Feynman diagrams made their first public appearance in a presentation of Richard P. Feynman at a small gathering of elite physicists at Pocono (Pennsylvania) in the spring of 1948 \cite[p. 43]{kaiser}. However, the reception was cold, presumably since this introduction was based on merely intuitive arguments (among them the unfamilar idea of path integrals) rather than supported by a mathematical basis which at that time did not exist \cite[p. 443]{schweber}. But this new diagrammatic technique captured the interest of Freeman Dyson, then a PhD student under Hans Bethe at Cornell where Feynman was staff member.\footnote{Dyson did not attend the Pocono meeting although Bethe had asked Oppenheimer (the host) to invite him. However, the meeting should be kept small and students -- no matter how bright -- were not considered \cite[p. 501]{schweber}.} Dyson managed to provide the derivation of the corresponding rules  from the mathematical basis of quantum electrodynamics (QED) and systematized the use of the diagrams significantly. He showed the equivalence of the three approaches by Julian Schwinger, Tomonaga Sin-itiro and Feynman. By combining the virtues of these approaches he  could even generalize the proof of renormalizability of  QED \citep{dyson1949}.\footnote{This paper by Dyson even preceded Feynman`s first work applying this technique \citep{feynman1949}. While Dyson`s work was published in February 1949 (submitted October the year before) Feynman's paper appeared in September 1949 (submitted in May 1949). The first Feynman diagram in print was therefore published by Dyson while occasionally the Feynman paper is quoted for ``the first published Feynman diagram'' \citep{wilczek2016}.} 

However, this technique did not spread on its own, especially since the first textbook treatments were available only after 1955 \cite[p. 27]{kaiser}. Instead it needed  what David Kaiser has called a ``postdoc cascade''.  Dyson moved to Princeton for his second year of his Commonwealth Fellowship and introduced a group of fellow postdocs at the Institute in the application of the method. These people took positions in academia scattered across (mainly) the US and developed schools which practiced the application of Feynman diagrams further.\footnote{Technically, Dyson was not part of this postdoc cascade, since he never received a PhD! After a year at the Institute Dyson moved back to the UK to work with Peierls in Birmingham. For administrative reasons he could not obtain an advanced degree there and was offered his professorship at Cornell in 1951 nonetheless \cite[p. 235]{peierls85}.}  

At the same time Feynman diagrams demonstrated a high plasticity, i.e. its use could be adopted to the needs of the local research groups and went soon beyond the original scope \cite[p. 173ff]{kaiser}. Initially designed for perturbative calculations within weakly coupled theories like quantum electrodynamics the diagrams soon found application in nuclear physics and meson theory. Here, only first order contributions could be classified with the help of the diagrams and higher order calculations were not even useful. 

The application in meson theory was championed by Robert Marshak and his group at Rochester. Marshak was certainly not part of the ``postdoc cascade'' but had attended the Pocono meeting.  Subsequently, he and his group benefited from frequent visits by Feynman over the next two years \cite[p. 221]{kaiser}. Thus, their use of the diagrams was more intuitive and framed into a particle language rather than the ``Dysonian'' style. Kaiser remarks at this point, that due to the   
\begin{quote}
``[...] failure of perturbative approaches for treating nuclear forces, Feynman diagrams' pictorial forms and calculational roles became more and more differentiated.'' \cite[p. 207]{kaiser}
\end{quote}

Now, the expression ``pictorial forms'' points already to the question whether Feynman diagrams could be used to picture the underlying process. 
 On this question Feynman and Dyson disagreed from the very beginning. As noted above it was Dyson who derived the ``Feynman rules'' for the translation of diagrams into the corresponding mathematical expression from a quantum field theoretical starting point. Before that he could not make any sense of them, or as he put it: 
\begin{quote}
``Nobody but Dick [Feynman] could use his theory, because he was always invoking his intuition to make up the rules of the game as he went along. Until the rules were codified and made mathematically precise, I could not call it a theory.'' (cited from \citet[p. 188]{kaiser})
\end{quote}
To Dyson, Feynman diagrams were essentially a bookkeeping device. In his lecture ``Advanced quantum mechanics'', delivered at Cornell during autumn 1951, he expressed this pointedly:
\begin{quote}
``We have introduced the Feynman graphs simply as a convenient pictorial way of visualizing the analysis of an operator into its normal constituents. The graphs are just diagrams drawn on the paper.''
\end{quote}
But he continues immediately: 
\begin{quote}
``But according to Feynman, ``Space-time Approach to Quantum Electrodynamics'', ({\em Phys. Rev.} {\bf 76} (1949) 769), the graphs are more than this. He regards the graphs as a picture of an actual process which is occurring physically in space-time.'' \cite[p. 99]{dyson1951}
\end{quote}

According to Dyson the diagrams sole purpose is to visualize the formulae. Thus, while the above mentioned  ``pictorial'' role and the separation from the mathematical underpinning  was foreign to Dyson, it was apparently closer to Feynman's thinking.\footnote{The origin of Feynman diagrams has been  investigated more closely by Adrian \cite{wuethrich2010}.} This marks the ``Feynman-Dyson split'' as David Kaiser has dubbed this 
\begin{quote}
``[...] tension between Feynman's and Dyson's positions, with their varying emphases on `intuition' versus derivation, physical pictures versus topological indicators.'' \cite[p. 263]{kaiser}
\end{quote}
 Kaiser shows how textbooks covering Feynman diagrams enhance this tension. While they typically announce a presentation in accordance with Dyson's rigorous derivation, they postpone this part until Feynman diagrams were introduced already -- apparently intending the diagrams to speak for themselves (compare \citet[Chap. 7]{kaiser}). This tradition continues until today. The excellent text by Matthew \citet{schwartz2014} for example introduces Feynman diagrams already in Sec. 4 while the derivation of the corresponding rules is postponed to Sec. 7.

It is of course the visual nature of Feynman diagrams which makes them a popular teaching tool and which is part of the explanation for their  great success. Kaiser argues that especially the resemblance with Minkowski diagrams and also bubble-chamber images allowed them to be
\begin{quote}
``[...] taught in ways that borrowed from more elementary skills that had already become second nature for most young physicists.'' \cite[p. 23]{kaiser}
\end{quote}
That is, the split which divided Feynman and Dyson widened in the course of this teaching tradition further. The resulting ambiguous or almost schizophrenic situation is exemplified by the following quote from a popular textbook first published in the late 1960s:
\begin{quote}
``Because of the unphysical properties of Feynman diagrams, many writers  do not give them any physical interpretation at all, but simply regard them as  a mnemonic device for writing down any term in the perturbation expansion.  However, the diagrams are so vividly `physical-looking', that it seems a bit  extreme to completely reject any sort of physical interpretation whatsoever. [...] 
Therefore, we shall here  adopt a compromise attitude, i.e., we will `talk about' the diagrams as if they  were physical, but remember that in reality they are only `apparently physical'  or  {`quasi-physical'}\,''. \cite[p. 88]{mattuck67} 
\end{quote}
What Richard Mattuck describes as a ``compromise attitude'' should confuse the readers very much and it is unclear why the language should not reflect the actual status of the denoted objects more closely.

\section{Interpretation of Feynman diagrams\label{int}}

As remarked above, textbook authors who introduce Feynman diagrams many pages before its mathematical basis is explained apparently assume that they can somehow ``speak for themselves''. Now, what do they say? A verbal description of them easily falls into the habit of giving a space-time or even causal account of the sequence of ``events'' pictured. Ubiquitous is talk about ``electrons and positrons meeting at a vertex'', ``up quarks transforming into down quarks upon emitting a W boson'' or ``Higgs particles decaying into top and anti-top quarks''.  At the same time many textbook authors provide also a word of caution, like ``one must not take this interpretation too literally'' \cite[p. 67]{mandlshaw}. However, what exactly ``too literally'' means is usually left open. 

In this section I show why this word of caution is more than justified, that is, why properties of  Feynman diagrams  are at odds with any representational or modeling function with regard to the underlying physical process. Until recently the literature on the interpretation of Feynman diagrams has been rather sparse.\footnote{The July-August issue ({\bf{26}}(4)) of  ``Perspectives on Science'' has been a special issue on Feynman diagrams. This volume appeared after the present paper had been submitted.} General discussions can be found in \citet{brown1996}, \citet{elkins2008} and  \citet{meynell2008}. Probably the most systematic treatment was given by \citet{kuhlmann2010}.\footnote{Kuhlmann takes the arguments against a realistic interpretation of Feynman diagrams to depend on an additional hidden premise, namely a ``substance ontology''. If, he argues, one moves to an ``process ontology'' instead  these arguments could be circumvented \cite[p. 125ff]{kuhlmann2010}. However, Kuhlmann admits that details of such an ontology are not worked out yet (p. 121).}  The specific aspect of virtual particles has captured the interest of more authors \citep{bunge70,redhead88,weingard,fox2008,arthur2012}. I not only combine and strengthen some of these arguments but  will also provide new ones. Especially by exploiting the notion of ``topological equivalence''  a strong case against any literal reading can be made. Before doing so I need to introduce some technical background and terminology.  

\subsection{Technical background}
\begin{figure}
	\centering
 	\includegraphics[width=0.45\textwidth]{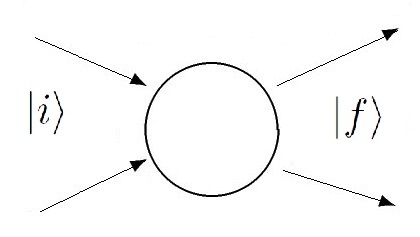}
	\caption{Schematic representation of a scattering process with two particles in the initial and final state. The circle in the center indicates the region of interaction.}
	\label{boi}
\end{figure}

Most of what we know about elementary particles comes from scattering experiments, i.e.  regards the probability for observing specific final states, given a specifically prepared initial state. For two particles in the initial and final state the situation is schematically depicted in Fig.~\ref{boi}. Here, the interaction is depicted as a circle which may be called the ``bubble of ignorance''. The operator which relates incoming and outgoing states is the so-called scattering matrix, formally defined as \cite[p. 81]{lahiri2000}:
\begin{eqnarray}
S = {\cal{T}} \left  [  \exp \left( -i \int_{-\infty}^{+\infty} d^4x {\cal{H}}_I \right ) \right ]. 
\end{eqnarray}
Here, ${\cal{T}}$ is the time-ordering operator and ${\cal{H}}_I$ the part of the Hamiltonian which describes the interaction. The range of integration indicates, that the interaction lasts only a finite time and that long before or long after the interaction the states are essentially free. The transition amplitude between such asymptotically free states $|i\rangle$ and $|f\rangle$ is now given by 
\begin{eqnarray}
\langle f|S|i\rangle = S_{fi}.
\end{eqnarray}

To calculate this amplitude  within QED or the Standard Model is very demanding. As a rule it can not be derived analytically and most predictions are obtained within perturbation theory, i.e. the solution is approximated by an expansion in powers of the coupling constant. Feynman diagrams are a tool to facilitate these calculations considerably.  Each element of the calculation can be graphically displayed  by a line or vertices between lines. Incoming and outgoing lines represent asymptotically free states and correspond to Dirac spinors  (fermions) or polarization vectors (photons) in the calculation. The lines are directed to distinguish between particles and antiparticles. Internal lines   correspond to a propagator or Green's function $G(x-x^{\prime})$ in the calculation of the $S$ matrix element. They represent ``virtual'' particles, i.e. states which do not have to obey the relativistic energy-momentum relation:
\begin{eqnarray}
E^2=p^2c^2+m^2c^4. \label{emr}
\end{eqnarray}
If they violate this relation they are  said to be ``off mass shell'' or simply ``off shell''. 

The points where lines meet are called ``vertices'' and contribute a vertex factor, $g$, to the calculation. Further more the 4-momentum is conserved at each vertex. Expressed technically, all diagrams to a given order $n$ represent a term $S_{fi}^{(n)}$  in the expansion 
\begin{eqnarray}
S_{fi}\approx S_{fi}^{(0)}+g S_{fi}^{(1)}+\cdots +g^N S_{fi}^{(N)}.
\end{eqnarray}
 Here,  $N$ denotes the order of the approximation, i.e. the highest power of the vertex factor $g$ entering. This factor $g$ is related to the coupling constant by $g \propto \sqrt{\alpha}$. In QED $\alpha$ is the fine structure constant, i.e. $g$ is proportional to the electron charge. If the spin is taken into account the vertex factor includes an additional term. In addition symmetry factors and sign conventions form part of the Feynman rules if applied to more complex situations. 

Now, the probability of the process is related to the square of the scattering matrix element:
\begin{eqnarray}
P_{i\rightarrow f} \propto |S_{fi}|^2\approx \left | S_{fi}^{(0)}+g S_{fi}^{(1)}+\cdots +g^N S_{fi}^{(N)} \right |^2. \label{square} 
\end{eqnarray}
To calculate a (differential) cross-section, i.e. the observable quantity,  one needs to include the information about the incoming flux and the available phase space of the corresponding process.

{We now turn to the interpretation of Feynman diagrams. To begin with I will review some of the arguments from the lierature. The reader familiar with this debate may jump directly to} Sec.~\ref{top} {where a novel argument is introduced.}

\subsection{The ``no-trajectory'' argument\label{no-t}}
On the most naive realistic reading of Feynman diagrams they are taken to depict actual particle trajectories. Why this attempt is misguided  is among  the oldest arguments against such a literal reading. Famously the uncertainty relations exclude the existence of   trajectories already in non-relativistic quantum mechanics. This was pointed out  on the first public presentation of Feynman diagrams in 1948 at the Pocono meeting by Niels Bohr \cite[p. 444]{schweber}. 

However, this is a rather weak argument since even Feynman in 1948 did not intend to picture particle trajectories and Niels Bohr acknowledged already at the Pocono meeting that this remark was based on a misconception \cite[p. 248]{mehra1994}. In fact, the existence of particle trajectories is no compulsory precondition  for  a realistic reading of Feynman diagrams (compare also \cite{wuethrich2012}). The following quote by one of the leading QFT experts and Nobel laureate Frank Wilczek is instructive:
\begin{quote}
``Feynman diagrams look to be pictures of processes that happen in space and time, and in a sense they are, but they should not be interpreted too literally. What they show are not rigid geometric trajectories, but more flexible, ``topological'' constructions, reflecting quantum uncertainty. In other words, you can be quite sloppy about the shape and configuration of the lines and squiggles, as long as you get the connections right.'' \citep{wilczek2016}
\end{quote}
Thus, by weakening the emphasis on the specific positions of the drawn elements compared to the overall ``topological'' structure an ``almost literal'' reading of the diagrams appears to be permitted. 

This remark presumably  alludes to the path integral approach to quantum theory. Famously, Feynman developed an alternative framework to quantum mechanics in which the probability amplitudes can be expressed as a sum (or rather functional integral) over all possible classical trajectories $x(t)$ weighted by the exponential of $i$ times the action ($e^{iS(x)/\hbar}$). This approach avoids the formalism of operators on Hilbert spaces. One may wonder how e.g. the uncertainty principle translates into this framework, given that it is usually associated with the non-commutativity of the corresponding operators. In the path integral approach the uncertainty relations reflect  quantum fluctuations of the classical variables $x(t)$ and $p(t)$ \cite[p. 927]{kleinert}. Presumably this is the root of Wilczek speaking about ``quantum uncertainty'' in terms of ``non-rigid trajectories''.

The path integral formalism can be generalized to cover (quantum) field theory. Again, the fields $\Phi(x)$ remain classical $c$-number fields and the quantum properties arise from the infinitely many integrals over $\Phi(x)$, one at each space time point $x$ \cite[p. 934]{kleinert}. While I will comment on the path integral approach to field quantization in some of the following sections, I should make the following general remark: The quantization via the path integral formalism has important applications (not only) in particle physics and may also provide additional insights into the underlying physics process. But since it  yields the same Feynman rules of perturbation theory it provides no difference for the interpretation of Feynman diagrams as discussed below.  

The above Wilczek quote mentions the ``topological'' aspects  of Feynman diagrams and in Sec.~\ref{top} I will come back to this issue. There I will argue that the underlying ``topological equivalence''  provides a powerful argument {\em against} a realistic reading of Feynman diagrams.

\subsection{The argument from superposition\label{sup}}
The innocent looking square in Eq.~\ref{square} (or actually the square of the modulus of this complex number) is key to clear up some confusions about the interpretation of Feynman diagrams. To interpret a {\em single term} of this sum physically ignores the effects which underlie already the double-slit experiment in quantum mechanics. There, in a very similar way, the {\em square} of the sum of the contributions of the different slits yields the observable interference pattern and the question whether the electron ``actually" went through slit 1 or 2 cannot be asked, let alone answered. In other words: the observed interference effects are related to {\em all} terms of the superposition. Conversely, a single term within a superposition does not refer to a physically realized process.

It might be objected, that there are cases in which a single diagram alone dominates the cross-section (see e.g. \citet{valente2011}). As noted by \citet[p. 128]{kuhlmann2010} it is common talk to call   higher order contributions `corrections'  to the `main process' which is given by the leading order term. I believe that also here a literal reading of Feynman diagrams can not be supported. For one thing, a general interpretation of Feynman diagrams can not be based on specific processes. In fact, there are many cases, in which the leading order calculation involves more than one Feynman diagram already. In addition, to view the physical process as being build up of independent subprocesses (i.e. the `main process' with additional `corrections')    produces false probabilities. On this view the observed probability should be given by the sum of the probabilities for each subprocess. Instead, each alleged ``subprocess'' contributes a term which has to be squared {\em after} summing to yield a probability. 

In order to illustrate this point further, consider a cross-section for which the leading order term alone can account for, say, 90\% of the observed value. This certainly does not support the claim that in 90\% of the cases this process actually proceeds with the exchange of a single virtual particle only, while only in the other 10\%   the higher order terms which involve loop corrections and the ``exchange'' of many particles occur. In fact the observed process is {\em always} described by the whole series.  

How this argument from superposition compromises the usual Higgs discovery narrative will be discussed in  Sec.~\ref{higgs-sup}.

In the literature the argument from superposition has been used to compromise the notion of ``virtual particles'', i.e. states depicted by the internal lines of Feynman diagrams  which do not have to obey  the relativistic energy-momentum constraint Eq.~\ref{emr} \citep{redhead88,weingard,fox2008}. However, if the internal lines are void of any real meaning, there is not much left for Feynman diagrams to represent, that is, these arguments translate naturally into points against a literal reading of Feynman diagrams {\em per se}. Conversely, on a literal reading of Feynman diagrams one is committed to grant some physical meaning to ``virtual''  states.

As noted by Robert \citet{weingard} one reason for taking virtual particle states seriously comes from the action of creation and annihilation operators on them. In fact, the {\em same} operators that change the particle number of external (``real'') particle states  act also on the internal states. Many textbooks on QFT  invoke the energy-time uncertainty to make the notion of virtual particles  plausible and explicitly claim that this relation implies the violation of energy conservation for the period $\Delta t \approx \hbar/\Delta E$.  This argument is disturbing for several reasons, as noted already by Mario \cite{bunge70}.  For one thing I do not know of any sound argument for the statement that in quantum mechanics energy is not conserved. Successful applications of the energy-time uncertainty relation deal e.g.  with the energy-spread of states (say, the natural line width) and their lifetime. 
 Now, according to the Feynman rules the constraint of energy-momentum {\em conservation} is met at each vertex. Thus, it is just the other way around: The internal lines representing ``off-shell'' states is a {\em consequence} of exactly this energy-momentum {\em conservation} -- rather than being implied by any violation of this principle. 

According to  Robert \cite{weingard} and Tobias \cite{fox2008} the generic superposition states render the {\em number} and {\em type} of these ``virtual particles'' indefinite. These authors therefore view virtual particles as a mere artifact of the specific solution technique, namely perturbation theory. This view is further supported by the fact that the recursive method to solve field equations with the help of Green's functions can also be applied for classical field equations. Also here the expansion can be depicted by ``Feynman diagrams'' with external and internal lines -- thus there is nothing  particularly ``quantum'' about them.  Further more, within other solution techniques like  lattice gauge theory no ``virtual particles'' emerge. Likewise, one may argue that ``virtual particles'' do not emerge in the path integral formulation of QFT, since no creation or annihilation operators enter the calculation of the Green's function \citep[p. 47ff]{weingard}. 

If one follows these arguments virtual particles should be viewed as fictitious. However,  within the sophisticated model-debate in the philosophy of physics it has been suggested that models may actually (and rightfully) contain ``fictitious'' elements. Alisa \citet{bokulich2009} has pointed to the useful distinction between ``explanatory fiction'' and ``mere fiction'' in this context.  Richard \cite{arthur2012}  applies this distinction to virtual particles and provides strong arguments for the claim that they are ``mere fiction'', i.e. serve no explanatory function. 

\subsection{The argument from ``topological equivalence''\label{top}}
This brings us to an argument which -- according to our knowledge -- has not been exploited in the debate so far. Suppose that somebody is not convinced by the above argument from superposition. For example  Mario Bacelar \cite{valente2011} argues in favor of virtual particles being more than a ``formal tool''. He considers cases which are dominated by single diagrams and tries to exploit the fact that the $S$ matrix expansion is only asymptotic, i.e. that an infinite superposition is not involved. 

However, what kind of ``story'' could a single Feynman diagram convey? Here one needs to recall an important concept from the theory of this diagrammatic technique.  Feynman diagrams which can be continuously deformed into each other while leaving the in- and out coming states unchanged are called ``topologically equivalent'' \cite[p. 68]{mandlshaw}. Topologically equivalent  diagrams depict the same amplitude and only one representative of this equivalence class  needs to be considered (presumably this concept details the remark about ``topological structures'' in the above Wilczek quote; see  Sec.~\ref{no-t}). 
\begin{figure}
	\centering
	\includegraphics[width=0.95\textwidth]{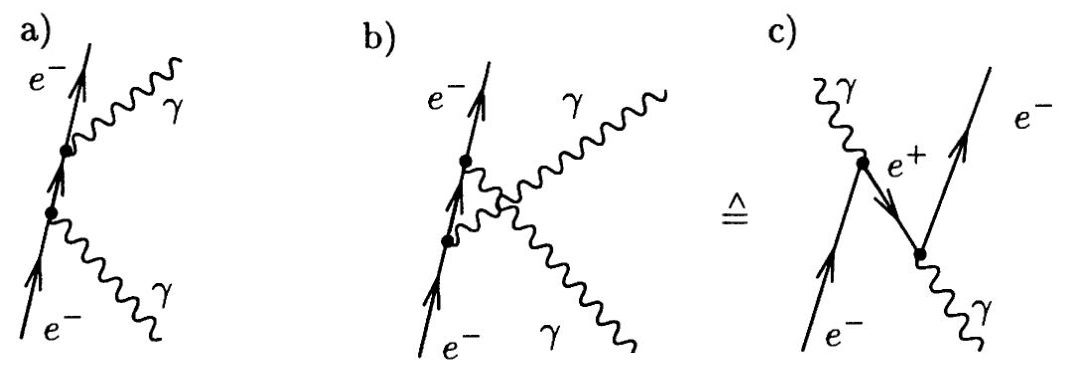}
	\caption{
Feynman diagrams for  Compton scattering in lowest order (time flow from bottom to top). In diagram a) a photon is first absorbed and subsequently emitted. The diagrams b) and c) differ only by time-ordering and are ``topological equivalent'', i.e. it is sufficient to consider just one of them.
	}
	\label{FD}
\end{figure}

This fact can be easily turned into an argument against the literal interpretation of Feynman diagrams which tries to read a ``sequence of events'' into this graphical representation. Let us look at the example of Feynman diagrams depicted in Fig.~\ref{FD} (time flow from bottom to top). 
They show the different contributions to the Compton scattering ($e^-\gamma\rightarrow e^-\gamma$) in lowest order perturbation theory. Fig.~\ref{FD} a) seems to capture the intuitive sequence of ``events'': first one photon is absorbed and subsequently one photon is emitted. However, already in lowest order another diagram needs to be considered as well. This is the one depicted in b) where photon emission and absorption reverse their time-order. This appears already counter-intuitive. However, an other -- topological equivalent -- diagram is shown in c). Here, the incoming photon splits into $e^+e^-$ while the incoming electron annihilates with the position to give rise to the final state photon. The electron of the pair creation is part of the final state. Thus, the stories which can be told to describe the diagrams b) and c) are quite different -- they even involve different types of particles. But given their equivalence only one needs to be considered. They represent only different ways to draw the {\em same} diagram, consequently, the Feynman graphs do not support a specific ``story'' to be told about the underlying process. 

In Sec.~\ref{higgs-topo} this argument will be applied to question the usual narrative of the Higgs discovery.


 \subsection{Summary of the interpretation of Feynman diagrams}
 

Summing up, there are major problems to sustain  the view that Feynman diagrams represent any underlying physical process. This, in particular, affects the notion of ``particle exchange'' as mediating interactions. Friebe et al.  suggest that this should be viewed as a metaphor only \cite[p. 242]{friebe2018}. For similar reasons James R. \citet[p. 265]{brown1996} concludes that Feynman diagrams ``do not picture any physical processes at all. Instead, they represent probabilities (actually, probability amplitudes)''. 
He compares them with Venn diagrams which also provide the visualization of an abstract relation (``being part of''), without claiming that the little circles resemble the corresponding system in any other respect. James Elkins comments in the same vein on the role of Feynman diagrams in the teaching of quantum field theory:
\begin{quote}
``Feynman diagrams are a useful calculational shorthand, but how helpful is a `learning tool' that mispresents its object so deeply and leads people to misuse it so persistently?'' \cite[p. 200]{elkins2008}
\end{quote}
He continues, that while Feynman diagrams no longer set out to resemble what they depict, the thinking about them is apparently influenced by a general attitude towards pictures:
\begin{quote}
``Yet, at the same time [...]  physicists, teachers, and students continue to think of them as if they could {\em also} have realistic properties -- as if to say every picture must have some realism simply to be a picture.''\cite[p. 200]{elkins2008}
\end{quote}
In this respect Feynman diagrams share features with other visualizations in quantum theory. To use an other example given by \cite{elkins2008}, the discrete spin values allowed by an electron are often depicted as vectors on a cone as if the electron would resemble a spinning top.

\section{The Higgs discovery as a case study\label{higgs}}
 Up to this point I have  strengthened the received view which takes Feynman diagrams as a mere paper tool without representational function proper. However, at the same time this view is in apparent tension with successful scientific practice. Let us look at a recent example, namely the discovery of the Higgs boson at the Large Hadron Collider (LHC) announced in 2012, in order to connect our abstract findings about Feynman diagrams  with the current research practice in high energy physics.  
\begin{figure}
	\centering
	\includegraphics[width=0.45\textwidth]{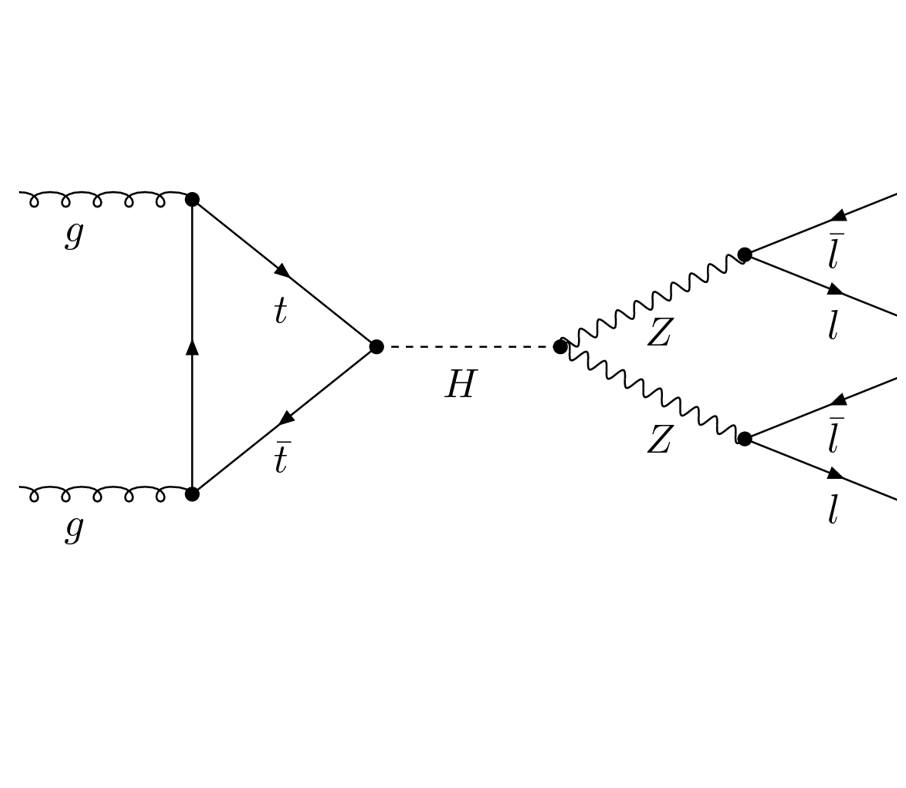}
	\includegraphics[width=0.48\textwidth]{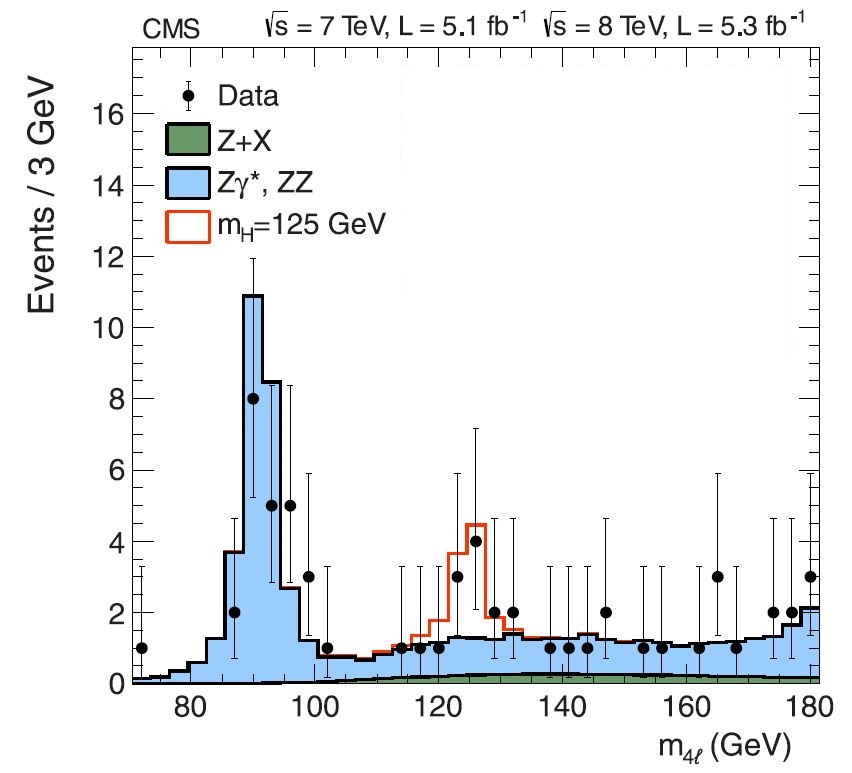}
	\caption{Left: Representative Feynman diagrams for the Higgs production via gluon-gluon fusion and decay into four charged leptons (time flow from left to right). Right: Distribution of the four-lepton invariant mass for the analysis of the $H\rightarrow ZZ\rightarrow 4\ell$ channel. The points represent the data, the filled histograms represent the background, and the open histogram shows the signal expectation for a Higgs boson of mass $m_{\mathrm{H}} = 125$ GeV, added to the background expectation.}
	\label{higgs4l}
\end{figure}

The dominant Higgs production mechanism at the LHC involves gluon-gluon fusion via an intermediate top-quark loop (see Fig.~\ref{higgs4l}, left). The Higgs discovery was based on the analysis of specific decay channels; most importantly: $H\rightarrow\gamma\gamma$, $H\rightarrow ZZ\rightarrow 4 \ell$ or $H\rightarrow WW \rightarrow 2\ell 2\nu$.  Here, $\ell$ denotes a charged lepton. The corresponding publications \citep{atlas,cms} devote a subsection to each decay channel, i.e. report the results of the different analyses separately (clearly, they are combined subsequently). So,  ATLAS and CMS have apparently ``observed'' the Higgs decay into two photons or four leptons separately. Contrary to what I said before, this seems to support the literal reading of  single Feynman diagrams like the one displayed in Fig.~\ref{higgs4l} (left).  

We should mention that the Higgs discovery publications by ATLAS and CMS do not contain Feynman diagrams but give reaction equations (as I did above). However, the announcement of the Higgs discovery by the CMS group in 2012 featured a very large number of Feynman diagrams. Also ATLAS used a few.\footnote{See  \url{https://indico.cern.ch/event/197461/} for the slides of the 2012 presentations}

\subsection{Higgs production}
Let us examine this specific production channel ($gg\rightarrow H$) more closely. Already in leading order there are two Feynman diagrams which contribute to the top-loop mediated process $gg\rightarrow H$  \citep{nikef}; not to mention the contributions from other quark flavors.\footnote{For gluon fusion the contributions from the bottom and the top quark have been taken into account by the LHC experiments. While the b-quark contribution to the cross section is only $\approx 1$\% the interference between these contributions has been found to be much larger (and destructive) \cite[p. 173]{wolf2015}.} The LHC experiments use the next-to-next-to-leading order (NNLO) prediction from QCD and apply NLO electroweak corrections. In addition the effects of so-called ``leading logs'' have been considered to {\em all} orders \cite[p. 2]{atlas}. Thus, to draw just a single diagram for the gluon-gluon fusion process $gg\rightarrow H$ is only a convenient shorthand for the whole partial series \cite[p. 53]{stoeltzner2017a}. To the expert all this is certainly well known. 


\begin{figure}
	\centering
	\includegraphics[width=0.35\textwidth]{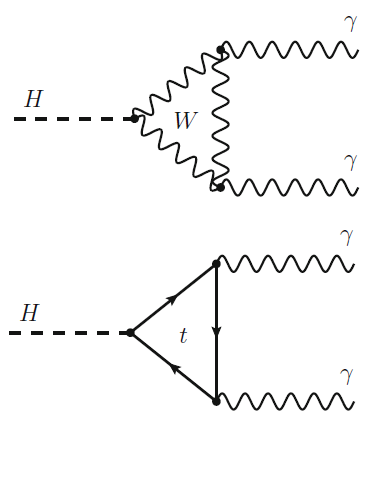}
		\includegraphics[width=0.6\textwidth]{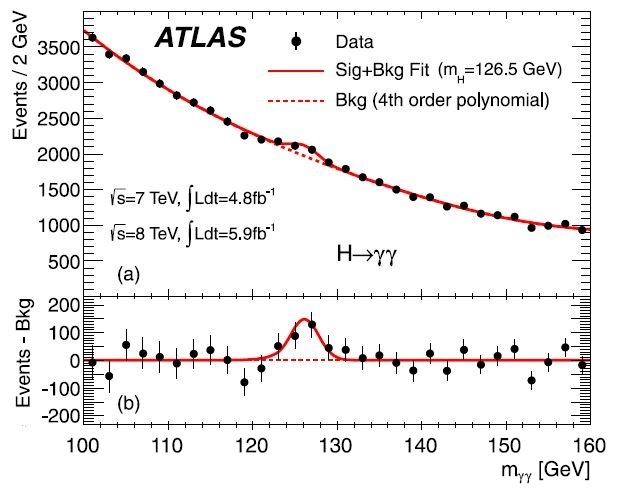}
 	\caption{Left: Leading-order contributions to the decay of the Higgs boson into two photons. Right: Two-photon Invariant mass distribution as measured by ATLAS in the Higgs discovery paper.}
	\label{hgg}
\end{figure}
\subsection{Higgs decay}
Let us move to the second part of the diagram, i.e. the Higgs decay channel $H\rightarrow ZZ\rightarrow 4\ell$ (in the LHC Higgs search the modes $\ell = e,\mu$  are considered). For one thing, if the final state contains identical leptons the corresponding interference effects need to be considered \cite[p. 2]{atlas}. Again, this effect can not be read off the Feynman diagram as depicted in Fig.~\ref{higgs4l} (left). In the case of $H\rightarrow \gamma\gamma$ there is more than one contributions in leading order. As depicted in Fig.~\ref{hgg} (left) the decay may either proceed via an intermediate $W$-loop or fermion-loop. Due to its huge mass the top quark yields the dominant contribution.  In cross section calculations the top loop enters with a minus sign, leading to a destructive interference term and lowering the overall decay rate into photons \cite[p. 93]{wolf2015}.

More importantly, one needs to ask how exactly this decay channel can be extracted from the data.  In order to claim a discovery an experiment needs to show that the probability for the data being produced by a fluctuation of the known background is less that $5.8\cdot 10^{-7}$ (this corresponds, in the Gaussian limit, to the famous $5\sigma$ significance criterion). Given that the cross-section for the Higgs production is tiny compared to processes with similar signatures this background needs to be understood with extreme precision.  In fact, the object of observation is not just the finally extracted signal, but the combination of many ``processes''.  In order to disentangle them one needs the knowledge of the relative fractions of all these cross-sections and has to normalize all contributions properly.  Again, the evaluation of countless diagrams (not to mention the correction due to detector- and hadronisation effects) contributes to the final result.

A typical result of this analysis, taken form the CMS Higgs discovery paper, is shown in  Fig.~\ref{higgs4l} (right). It shows the four-lepton invariant mass as reconstructed from the selected data \citep{cms}. Most prominent is the $Z$ peak arround 91 GeV, which is due to resonant $Z$ production.\footnote{We note in passing that this peak is very useful for the energy  calibration -- something missing in the invariant mass distribution of the $\gamma\gamma$ events.} In the range of 120 and 130 GeV an excess of events above the expected background can be seen. The open histogram corresponds to the Standard Model prediction with a Higgs mass of $m_{H}=125$ GeV and fits the data well.  Only something like 15 events contribute to this Higgs signal. It is evident that in order to derive any result the background needs to be controlled with extreme  precision.

Thus, it would be misleading to claim, that the analysis of the $H\rightarrow ZZ\rightarrow 4\ell$ channel just ``measures'' or ``observes'' the contribution from a single Feynman diagram; even if the signal as such is dominated by the contribution of only one or a few Feynman graphs indeed. 

Agreed, a proper understanding of the background, calibration etc. is not specific to LHC physics. And indeed, an invariant mass distribution as the one in Fig.~\ref{higgs4l} (right) or the corresponding distribution for the decay channel $H\rightarrow\gamma\gamma$ (Fig.~\ref{hgg}, right)  seems to suggest that the underlying process is  captured by ``decay channel talk''. Apparently such a distribution is as close as one can get to actually ``see'' a Higgs boson via the corresponding decay. However, there is an additional twist to the argument. Suppose that the invariant mass distributions could be measured with highly improved accuracy. Would this distributions peak at the ``actual'' Higgs mass? The surprising (and to our mind not sufficiently emphasized) answer is no, as will be explained in the next subsection.

\subsection{How to extract the Higgs mass -- the superposition argument in the Higgs decay\label{higgs-sup}}
\begin{figure}
	\centering
	\includegraphics[width=0.5\textwidth]{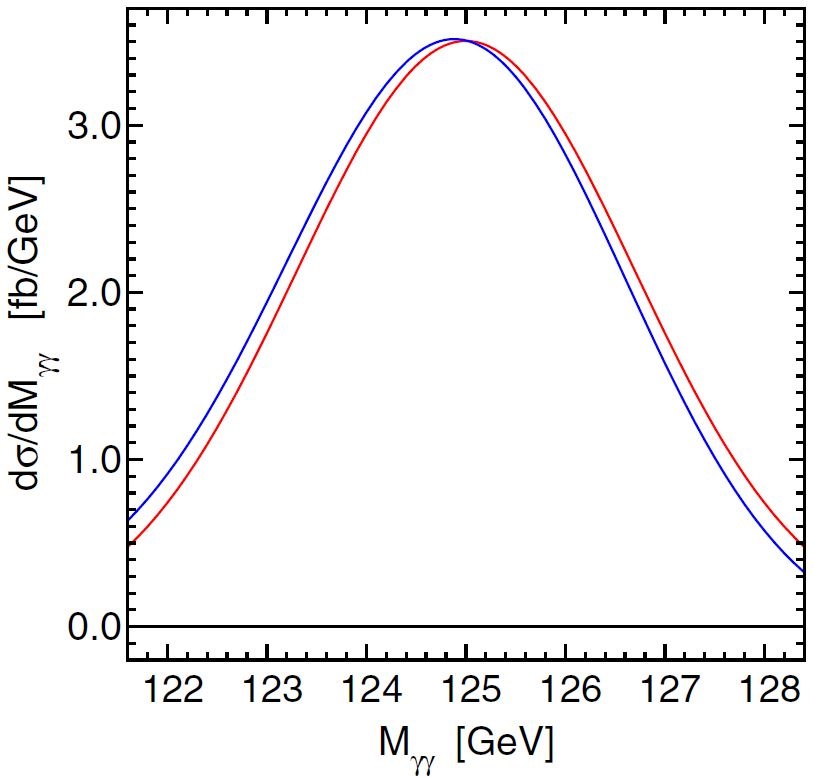}
	\includegraphics[width=0.45\textwidth]{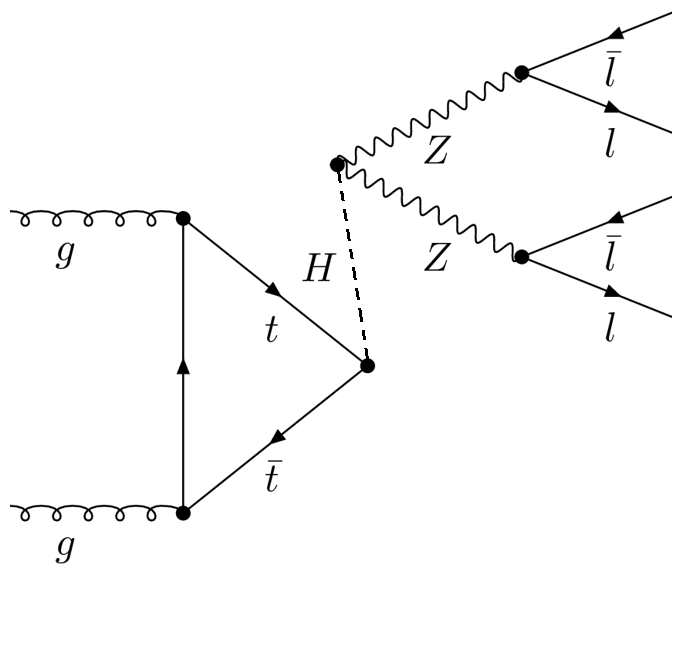}
	\caption{Left: Mass shift of the 2-photon invariant mass distribution due to interference effects with the background. The right curve shows the Higgs contribution only, while the mass peak is shifted to lower values if the interference effects are taken into account (Fig. taken from \citet[Fig. 3]{martin2012}). Right: Topological equivalent Feynman diagram for one of the  amplitudes which yield the dominant contribution in the Higgs discovery (time flow from left to right).}
	\label{topo}
\end{figure}

If the invariant mass distribution in the $\gamma\gamma$ or $4\ell$ channel could be measured with highly improved accuracy the peak of the distribution would not correspond to the Higgs mass, since  due to interference effects with the background the distribution gets shifted. This effect has been discussed for the channel $H\rightarrow ZZ$ by \cite{kauer2012} and for the $H\rightarrow \gamma\gamma$ channel by \cite{martin2012}. This last paper contains also the plot reproduced in Fig.~\ref{topo} (left). Here, the right curve shows the Higgs contribution only, while the mass peak is shifted to lower values if the interference effects are taken into account. That is, in principle the Higgs mass can not be measured by directly fitting the mass-distribution in the corresponding plot, but only  by comparing the full Standard Model prediction (with varying $m_{H}$ values) with the obtained distribution. Strictly speaking, the peak in the invariant mass distribution is not identical to the Higgs mass and the ``process'' $H\rightarrow\gamma\gamma$ can not be observed in isolation. 

We agree that the shift in the mass distribution is rather small (estimated to $\approx100$ MeV   in the $\gamma\gamma$ channel \cite[p. 4]{martin2012} and even smaller for the Higgs decaying into $ZZ$). The combined Higgs mass analysis of ATLAS and CMS neglects background interference although the result yields $m_{H}=125.09\pm 0.21 (\mathrm{stat}) \pm 0.11 (\mathrm{sys})$ GeV \citep{atlascms2015}. Actually an effect in the order of some 10 MeV should be a sizable contribution to the systematic uncertainty.\footnote{Interestingly, it has been suggested that this interference could be   exploited to bound (or even measure) the Higgs width by ``interferometry'' \citep{dixon2013}. Thus, this at first undesired effect can be turned into a useful tool.} 

It goes without saying that all this is very well known in the community.  The interference effect with the background are dealt with in detail in the yellow reports of the {\em LHC Higgs Cross Section Working Group} (see e.g. \citet[p. 35ff]{higgswg}). In the first volume of the Higgs working group report  it is stated explicitly:
\begin{quote}
``[...] what the experiments observe in the final state is not always directly connected to a well defined theoretical quantity. We have to take into account the acceptance of the detector, the definition of {\em signal}, the interference {\em signal-background}, and all sorts of approximations built into the Monte Carlo codes. As an example at LEP, the line shape of the $Z$ for the final state with two electrons has to be extracted from the cross section of the process ($e^+e^-\rightarrow e^+e^-$), after having subtracted the contribution of the photon and the interference between the photon and the $Z$.'' \cite{dittmaier}[p. 2] (emphasis in the original)
\end{quote}

In any event I take this numerically small effect to be of great conceptual importance. 
The interference between Higgs and background processes  demonstrates that the whole talk about ``Higgs decay channels'' holds only approximately. The term ``channel'' suggests that all events can be grouped into disjoint classes, while actually  interference takes place between signal and background contributions if the final states are the same. Given that the  dominant contribution from resonant Higgs decays comes from the  Higgs propagator the neglect of these interference effects are a justified  approximation but do not challenge our previous claim about Feynman diagrams. The argument based on the superposition of amplitudes (see Sec.~\ref{sup}) remains valid since it does not presuppose that the corresponding effects are {\em large} -- it just assumes that these effects are {\em possible}. Consequently, this valid numerical approximation does not support the intuitive picture of Higgs production and subsequent decay which follows from a literal reading of the Feynman diagrams. This narrative singles out  specific (signal) amplitudes while the actual process is related to the squared sum of {signal and background contributions}. Thus, one is not talking about ``ordinary'' background which could be eliminated by applying more clever cuts or by subtraction. For these kinds of contributions the term ``irreducible background'' has been coined. The corresponding invariant mass plots show a Higgs signal, to be sure. Its mass can be deduced from it.  But these plots do {\em not} display the Higgs mass.

It should be noted that ATLAS and CMS do not identify the mass peak with the Higgs boson explicitly. Instead, they call the corresponding region in their histograms ``signal expectation'' or ``signal component'' for a Higgs with $m_H=125$ GeV \citep{atlas,cms}. However, as indicated above, also the separation into ``signal'' and ``background'' holds only approximately. For example the background in the $\gamma\gamma$ analysis was fitted by a polynomial and subtracted from the data to yield the lower inset of the plot in Fig.~\ref{hgg} (right).  

\subsection{The argument from ``topological equivalence'' in the Higgs discovery\label{higgs-topo}}
We have seen above how the argument from superposition compromises the literal reading of Feynman diagrams in the case of Higgs production and decay. The same is true for the   argument from ``topological equivalence'' (see Sec.~\ref{top}).  Fig.~\ref{topo} (right) shows  a diagram which is equivalent to the one in Fig.~\ref{higgs4l} a). Here, the time-ordering of Higgs creation and annihilation is reversed. Under a literal understanding of Feynman diagrams the discovery would then be based on the Higgs production together with a pair of $Z$ bosons while the Higgs annihilates subsequently with a $t\bar{t}$-pair. Again, that is not the kind of story one wants to tell when dealing with a ``Higgs production in the gluon-gluon channel'' since it apparently deals with the top-loop induced Higgs {\em annihilation} instead. 

\begin{figure}
	\centering
	\includegraphics[width=0.5\textwidth]{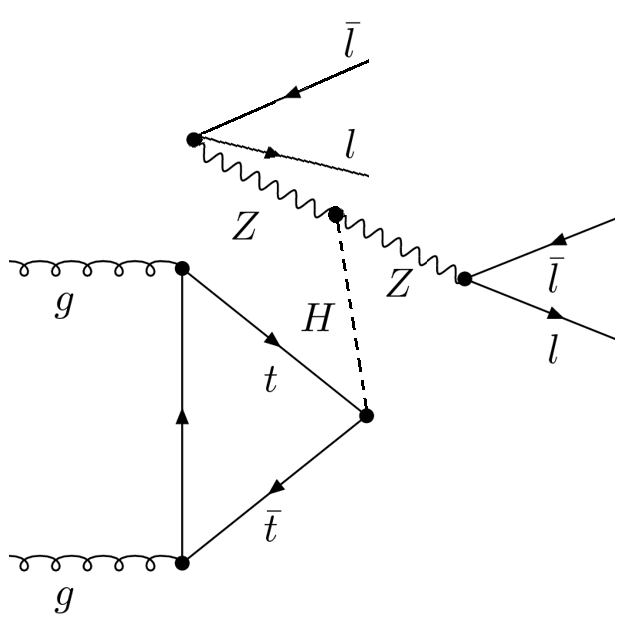}
	\caption{An other example for a topological equivalent Feynman diagram in the $H\rightarrow ZZ\rightarrow 4\ell$ channel.}
	\label{topo2}
\end{figure}

Of course you may invent still other equivalent diagrams for the same amplitude. Fig.~\ref{topo2} gives an example where the $Z$-line was twisted away. Here the apparent ``story'' would be the production of lepton pairs together with a $Z$ which subsequently radiates off a Higgs boson. The $Z$ produces another pair of leptons in the final state while the Higgs annihilates again with top-pairs. 
Also here, the literal reading of Feynman diagrams can not be sustained and the inside of the ``bubble of ignorance'' remains unknown.

\section{Summary and conclusion\label{sum}}

We have argued that  Feynman diagrams do not model or represent the underlying physical processes. While the often quoted ``no-trajectory'' argument may be circumvented, the argument from superposition questions the possibility to assign any meaning to a single Feynman diagram. However, even if individual diagrams could be interpreted physically, they would not provide a unique ``sequence of events''. The notion of topological equivalence compromises a literal understanding of these ``stick-figures'' \citep{kaiser2000} further.  Thus,  Feynman diagrams picture formulae -- not physical processes.

While all this just strengthens the received view I took issue with the apparent tension between this anti-realistic reading of Feynman diagrams and their use (and the common jargon) in high energy physics. For example in the Higgs discovery the common narrative appears to take Feynman diagrams as direct visualizations of a Higgs decaying into two photons or four leptons.  The division into these distinct decay channels suggests disjoint classes in which the events can be grouped. Furthermore,  a peak at $\approx 125$ GeV in the   invariant mass spectrum  apparently provides direct evidence for the  corresponding Higgs decay channel to be observed.

However, an individual event can not be assigned to the ``background'' or ``signal'' category. If the final states are the same the observed process is the superposition of the contributions from the corresponding diagrams and interference occurs. This shows up for example in the invariant mass spectrum which does not peak (exactly) at the Higgs mass value. Likewise, the argument from topological equivalence can be applied to compromise an intuitive understanding of Higgs production and subsequent decay. 

As expected, the   tension between the anti-realistic reading of Feynman diagrams and the successful research practice is only apparent. Of course, the material presented in our Higgs case study is well known (not only) to the experts in the field. The reason why all these aspects do not figure prominently in the Higgs discovery  lies in the fact that in resonant production the corresponding interference effects are {\em numerically} small and can not be experimentally resolved  yet.  It goes without saying that the Higgs analyses of ATLAS and CMS are admirable achievements and that the confirmation of the Higgs mechanism is an intellectual and technical masterpiece. The applied approximations are valid and I certainly do not question the discovery of the Higgs boson. However, the corresponding narrative  promotes (unintentionally, I assume) a simplistic picture of particle physics in which a literal reading of Feynman diagrams is supported. In (non-relativistic) quantum mechanics  it is usually a central claim that no causal or space-time picture of the process between preparation and measurement can be given. It is curious to note that on a literal reading of Feynman diagrams some authors apparently accept the risk  to fall  below this level.

Most importantly, the different amplitudes should not be confused with ``physical processes'', given that the actual observation relates to the squared sum of all contributions. In this sense a contribution like  $H\rightarrow \gamma\gamma$ can not be observed since this very category is not applicable. What has been successfully tested was the whole of the Standard Model including its account for electro-weak symmetry breaking.


As discussed in Sec.~\ref{history} David Kaiser has dubbed the tension between a more intuitive understanding versus the formal derivation of the diagrams  the ``Feynman-Dyson split''. He has noted that textbook presentations have enhanced this tension. Our case study reveals that the presentation of the current research practice in high energy physics  likewise contributes to a widening of this split.

\section*{Acknowledgment}
We thank Robert Harlander, Tilman Plehn and Thomas Z{\"u}gge for illuminating discussions and helpful comments. The valueable suggestions by the referees and the editors are gratefully acknowledge as well as the proof reading of Joan P. Marler.


\end{document}